\documentclass{amsart}
\usepackage{graphicx}
\usepackage{amssymb}

\newtheorem{problem}{Problem}
\newtheorem{theorem}{Theorem}

\newtheorem{prop}{Proposition}

\newcommand{\vc}[1]{{\boldsymbol{#1}}}
\newcommand{\R}{\mathbb{R}}
\newcommand{\U}{{\mathcal{U}}}
\newcommand{\eq}{\triangleq}

\begin{document}

\title[Sparsity Methods for Networked Control]{Sparsity Methods for Networked Control
\footnote{Submitted to IEICE SmartCom2014}.}
\author[M. Nagahara]{Masaaki Nagahara}
\address{M.~Nagahara is with 
Graduate School of Informatics, Kyoto University,
Kyoto 606-8501, JAPAN. (nagahara@ieee.org).}

\keywords{
networked control, sparsity, optimal control, hands-off control%
}

\maketitle

\begin{abstract}
In this presentation, we introduce sparsity methods for networked control systems 
and show the effectiveness of sparse control. In networked control, 
efficient data transmission is important since transmission delay and error can critically 
deteriorate the stability and performance. We will show that this problem is solved by sparse control 
designed by recent sparse optimization methods.
\end{abstract}

\section{Introduction}
Sparsity methods, called \emph{compressed sensing} or
\emph{sparse representation}, have been recently introduced
in signal processing \cite{EldKut}.
The methods are also applied to communications, see
a survey paper \cite{HayNagTan13}.

A sparse vector is a vector that
has very few non-zero entries compared with the vector size.
Compressed sensing takes advantage of sparsity property of signals in some domain
(e.g. the Fourier domain), for efficiently reconstructing such signals from very few measurements by
finding a solution of underdetermined linear equations.
To solve such underdetermined linear equations, one can adopt an $L^1$-norm minimization
\cite{CheDonSau98}
to achieve the sparsity, or a matching pursuit to find a sparse solution in a greedy way
\cite{MalZha93}.

Sparsity methods are very recently applied to solving problems in control systems
such as predictive networked control \cite{NagQue11,NagQueOst14},
hands-off control \cite{NagQueNes13},
actuator scheduling \cite{AguDelDolAgu14}, and
security in cyber-physical systems \cite{FawTabDig14},
to name a few.
In this presentation, we introduce sparsity methods for networked control systems
and discuss the effectiveness of the sparse control.

\section{Problem Formulation}
\label{sec:problem}
We here consider linear and time-invariant (LTI) plant models of the form
\begin{equation}
 \frac{d\vc{x}(t)}{dt} = A\vc{x}(t) + B\vc{u}(t), \quad t\in[0,T],
 \label{eq:plant}
\end{equation}
where
$\vc{x}(t)=[x_1(t),\ldots,x_n(t)]^\top\in\R^n$ is the state,
$\vc{u}(t)=[u_1(t),\ldots,u_m(t)]^\top\in\R^m$ is the control input,
and $T\in (0,\infty)$ is the length of the control horizon.

The control $\{\vc{u}(t): t\in[0,T]\}$ is chosen to drive the state $\vc{x}(t)$
from a given initial state 
$\vc{x}(0)=\vc{x}_0$
to the origin in time $T$, that is,
$\vc{x}(T)=\vc{0}$.
Also, the control $\vc{u}(t)$ is constrained in magnitude by
\begin{equation}
 \|\vc{u}(t)\|_\infty \leq 1,\quad \forall t \in [0,T].
 \label{eq:input_constraint}
\end{equation}
We call a control $\{\vc{u}(t): t\in[0,T]\}$ \emph{admissible}
if it satisfies \eqref{eq:input_constraint}
and the resultant state $\vc{x}(t)$ from \eqref{eq:plant} satisfies boundary conditions
$\vc{x}(0)=\vc{x}_0$ and $\vc{x}(T)=\vc{0}$.
We denote by $\U$ the set of all admissible controls.
Among all admissible controls in the set $\U$, we consider a control that
maximizes the time interval over which the control $\vc{u}(t)$ is exactly zero.
Such a control is called a \emph{maximum hands-off control},
the problem of which is described as follows:
\begin{problem}[Maximum Hands-Off Control]
\label{prob:MHO}
Find an admissible control $\{\vc{u}(t): t\in[0,T]\}\in\U$ that minimizes
\begin{equation}
 J_0(\vc{u}) \eq \sum_{i=1}^m\lambda_i \|u_i\|_{L^0} = \sum_{i=1}^m\lambda_i \int_0^T \phi(u_i(t))dt,
 \label{eq:J_MHO}
\end{equation}
where $\lambda_1>0,\dots,\lambda_m>0$ are given weights and
\[
 \phi(u) \eq \begin{cases} 1, & \text{if}~ u \neq 0,\\ 0, & \text{if}~ u = 0. \end{cases}
\] 
\end{problem}
This problem is quite hard to solve since the cost function
is highly nonlinear and non-convex.
To overcome the non-convexity, we introduce the following
$L^1$-optimal control problem:
\begin{problem}[$L^1$-Optimal Control]
\label{prob:L1}
Find an admissible control $\{\vc{u}(t): t\in[0,T]\}\in\U$ that minimizes
\begin{equation}
 J_1(\vc{u}) \eq \sum_{i=1}^m\lambda_i \|u_i\|_{L^1} = \sum_{i=1}^m\lambda_i \int_0^T |u_i(t)|dt.
 \label{eq:J_L1}
\end{equation}
\end{problem}
This problem is a classical $L^1$ optimal control problem,
also known as fuel-optimal control problem, and 
can be easily solved \cite[Chap.~8]{AthFal}.
Moreover, the following theorem shows the equivalence
between the two control problems
\cite{NagQueNes13}:
\begin{theorem}
\label{thm:L1}
Assume that $(A,B)$ is controllable%
\footnote{For the definition of controllability, 
see \cite[Sect.~4-15]{AthFal}. $(A,B)$ is controllable iff
$\mathrm{rank}[B,AB,\dots,BA^{n-1}]=n$.}.
Assume also that Problem \ref{prob:L1} has at least one solution%
\footnote{A linear system that is controllable and with nonsingular $A$
is called a \emph{metanormal} system \cite{Haj79}.}.
Then the set of the solutions of Problem \ref{prob:MHO}
(maximum hands-off control)
is equivalent to the set of the solutions of Problem \ref{prob:L1}
($L^1$-optimal control).
\end{theorem}
\section{Networked Control}
\label{sec:NC}
Let us assume that
the conditions in Theorem \ref{thm:L1} hold.
Then, the maximum hands-off control takes only 
3 values, $\{-1,0,1\}$, and the value changes discontinuously.
This property, called ``bang-bang control,'' benefits networked control systems
since the control value can be represented in only 2 bits.
Moreover, the number of switching times is bounded as shown in the following
proposition:
\begin{prop}
\label{prop:switching}
Assume that $(A,B)$ is controllable and
$A$ is nonsingular.
Let $\omega$ be the largest imaginary part of the eigenvalues of $A$.
Then, the maximum hands-off control is a piecewise constant signal,
with values $-1$, $0$, and $1$,
with no switches from $+1$ to $-1$ or $-1$ to $+1$,
and with $2nm(1+T\omega/\pi)$ discontinuities at most.
\end{prop}
{\bf Proof:} 
Theorem \ref{thm:L1} combined with
Theorem 3.2 of \cite{Haj79} gives the results.
\hfill$\Box$

Let us consider a networked system where
we should send the control signal $\vc{u}(t)$ on time interval $[kT,(k+1)T)$
at every sampling time $kT$, $k=0,1,2,\dots$.
From Proposition \ref{prop:switching},
we use 1 bit for representing the change of the control values,
and $b$ bits for representing each switching time.
From Theorem \ref{prop:switching}, we need in total
\[
 1 + 2nmb\left(1+\frac{T\omega}{\pi}\right) ~ \text{[bit]}
\]
to represent the maximum hands-off control on time interval $[0,T]$,
or
\[
 \frac{1}{T} + 2nmb \left(\frac{1}{T}+\frac{\omega}{\pi}\right) ~ \text{[bps]},
\]
which is much smaller than representing a general signal on $[0,T]$.
This is an advantage of the maximum hands-off control for networked control systems.

\section{Conclusion}
In this presentation, we have introduced maximum hands-off control (the sparsest control)
and shown that this control is equivalent to $L^1$-optimal control under some assumptions
on the optimal control problem.
The maximum hands-off control has a ``bang-bang'' property, which is very
advantageous to networked control systems in view of compressed data representation.

\end{document}